SpaceOps-2025, ID # 407

## Lights-Out: An Automated Ground Segment for unstaffed Satellite Operations

### Marvin Böcker[a]*, Ralph Biggins[b], Michael Schmeing[c]

[a] *CGI Deutschland B.V. & Co. KG, Alte Wittener Straße 56, 44803 Bochum, Germany,* *marvin.boecker@cgi.com*
[b] *CGI Deutschland B.V. & Co. KG, Mornewegstraße 30, 64293 Darmstadt, Germany,* *ralph.biggins@cgi.com*
[c] *CGI Deutschland B.V. & Co. KG, Alte Wittener Straße 56, 44803 Bochum, Germany,* *michael.schmeing@cgi.com*
\* Corresponding Author

### Abstract

We present our approach for a periodically unstaffed, fully automated ground segment. The concept is in use for the first time on the German satellite communications mission Heinrich Hertz on behalf of the German Space Agency at DLR. Heinrich Hertz was launched in July 2023 and offers access to scientific and technical experiments to its users.

The mission utilizes major automation concepts for the satellite platform operations, allowing fully automated operations outside of office hours. The concept includes tracking, telemetry and commanding (TTC) of the satellite. Pre-planned and automatically executed schedules enable commanding without human interaction. The user mission schedule is planned separately from the main mission schedule and is automatically de-conflicted. The automatic monitoring concept monitors the systems of the satellite and all assets in the ground segment and triggers reactions in operator-configurable ways depending on the mission needs, for example emergency notifications or automated execution of flight operation procedures.

Additionally, the concept also puts special emphasis on a self-service user portal that provides flexible access 24/7, even when the control center is not staffed. The portal allows external users of the payload to schedule pre-defined experiments, monitor the live execution of the experiment with browser-based displays and access ground station telemetry and dedicated RF test equipment during the time of their scheduled experiment. Tasks can be planned long in advance as well as with a short reaction time (less than 1 minute), which allows, for example, the reconfiguration of the payload during a running experiment.

In this paper, we describe the automation approach to the operational concept in detail and discuss the transition from LEOP to lights-out mode. We also include feedback from external experimenters about their experience with the self-service portal and the options the ground segment gives them in terms of monitoring and control of their payload.

**Nomenclature**
- External User – Scientific/Technical User or Military User
- W/T User / Experimenter – Synonymous for Scientific/Technical Users

**Acronyms/Abbreviations**
- AR – Activity Request
- CSM – Centralized Signal Monitoring
- FOP – Flight Operation Procedure
- GOP – Ground Operation Procedure
- H2Sat – Heinrich Hertz Satellite Mission
- HMI – Human Machine Interface
- MIB - Mission Information Base
- MIL – Military
- SCS – Satellite Control Segment
- UER – User Execution Request
- VS-NfD – Classified information - for official use only
- W/T – wissenschaftlich/technisch (scientific/technical)

Internal

## 1. Introduction

Since a few years, the requirements for satellite missions and the associated ground segments have been undergoing a strong change. Concepts such as unstaffed and automated control centers, which can be flexibly scaled, are becoming more and more important. This paper presents a new approach for a highly automated control center that is periodically unstaffed. The concept is used for the first time on the German communications satellite named Heinrich Hertz (also called "H2Sat"). This mission is led by the the German Space Agency at the German Aerospace Center (DLR) in Bonn on behalf of the Federal Ministry of Economics and Climate Protection (BMWK) and with the participation of the Federal Ministry of Defence (BMVg). On July 6th, 2023, the Heinrich Hertz mission became the first German communications satellite to be launched into space on the last European Ariane 5 rocket to research and test new technologies and communications scenarios. The mission offers universities, research institutes and industry a platform to conduct In-Orbit Verification of novel platform and payload technologies as well as a variety of scientific and technical experiments in the field of new communication technologies. Apart from the scientific and technological part of the mission, the cooperation with the German Federal Ministry of Defence (BMVg) provides additional and independent communications for the German armed forces (Bundeswehr).

The mission also aims to prove new technologies and German industry capabilities like advanced monitoring and control technologies on ground. There are two main goals for automation in Heinrich Hertz: (a) lights-out operations (platform and payload) and (b) external user self-service (payload) using a secured web UI.

The new concept for ground segments presented in this paper shows technologies for the periodically fully automated operation of an entire ground segment. The control centers are only staffed during normal office hours from 8am to 5pm from Monday to Friday ("lights-out operations"). The concept includes tracking, telemetry and commanding (TTC) of the satellite as well as 24/7 automated support for user missions. Pre-planned and automatically executed schedules enable commanding without human interaction. The user mission schedule is planned separately from the main mission schedule. Through automated conflict resolution, the schedules are combined and prepared for automated processing. This concept allows a flexible scaling of the scientific users and of the connected user segments. Thus, it is possible to add a new user segment with any number of users connecting to the main mission control at any time.

While this paper covers the technical aspects of lights-out operations and external user access, there are two more papers covering Heinrich Hertz at Space Ops 2025: The paper by A. Moorhouse et al. [1] covers the general mission of the satellite from an operations aspect while M. Schmeing et al. [2] goes more in-depth into the planning software ("Pleniter Plan") and its advanced constraint solving algorithms, without focusing on the Heinrich Hertz mission.

## 2. Ground Segment Overview

The ground segment system that was designed and implemented for this mission consists of:

- Control centres (primary and backup) including all related systems necessary for manual and automated operations of the satellite and its payloads.
- Ground stations (primary and backup) including all required antennas, tracking systems, and equipment to facilitate communication with the H2Sat satellite. These stations are strategically located to maximize coverage and signal quality.
- Ground networks interconnecting all ground assets to enable reliable data exchange, coordination, and integration. The networks are designed with robust and BSI-certified security protocols to protect mission-critical operations.

CGI Deutschland B.V. & Co. KG was responsible to design, implement and deliver the ground segment for the H2Sat mission. Operations of H2Sat satellite since LEOP (Launch and Early Orbit Phase) have been carried out with CGI Pleniter® software. Pleniter supports all operational tasks of the H2SAT missions like mission analysis, operations preparation, system validation, LEOP, IOT (In Orbit Testing). It was tailored to the mission specific needs and grows along with the new mission requirements. In addition to these major functional blocks, the ground segment also contains interfaces to the different stakeholders and service providers to the mission.

In terms of infrastructure, the H2Sat Ground Segment consists of two control centres, a primary and a backup control centre, two antenna sites (both hosting TT&R and Anchor Antennas) as well as the premises of the external users. The two independent control centres are set for redundancy and contingency purposes, one acting as prime control centre and one as backup control centre. Each control centre is fully capable of operating the H2Sat satellite with the backup control centre becoming the temporary primary control centre. A failover to the backup centre is

Internal

triggered in case of a major anomaly on the prime. Both control centres have been established separated by a sufficient geographic distance to ensure that in case of fire, flooding or power failures they are not both affected simultaneously.

On a logical level, the H2Sat Ground Segment consists of two segments: The Satellite Control Segment (classified) and the W/T User Segment (unclassified). Please also see Figure 1.

- The Satellite Control Segment covers all facilities necessary to operate the satellite platform as such. This includes parts of the software system located in the Satellite Control Centre as well as the TT&R antennas located at the two ground station sites.
- The W/T User Segment comprises all components necessary to operate the W/T user mission. This includes parts of the software system located in the Satellite Control Centre, the anchor antennas located at the two ground station sites as well as the W/T User premises.

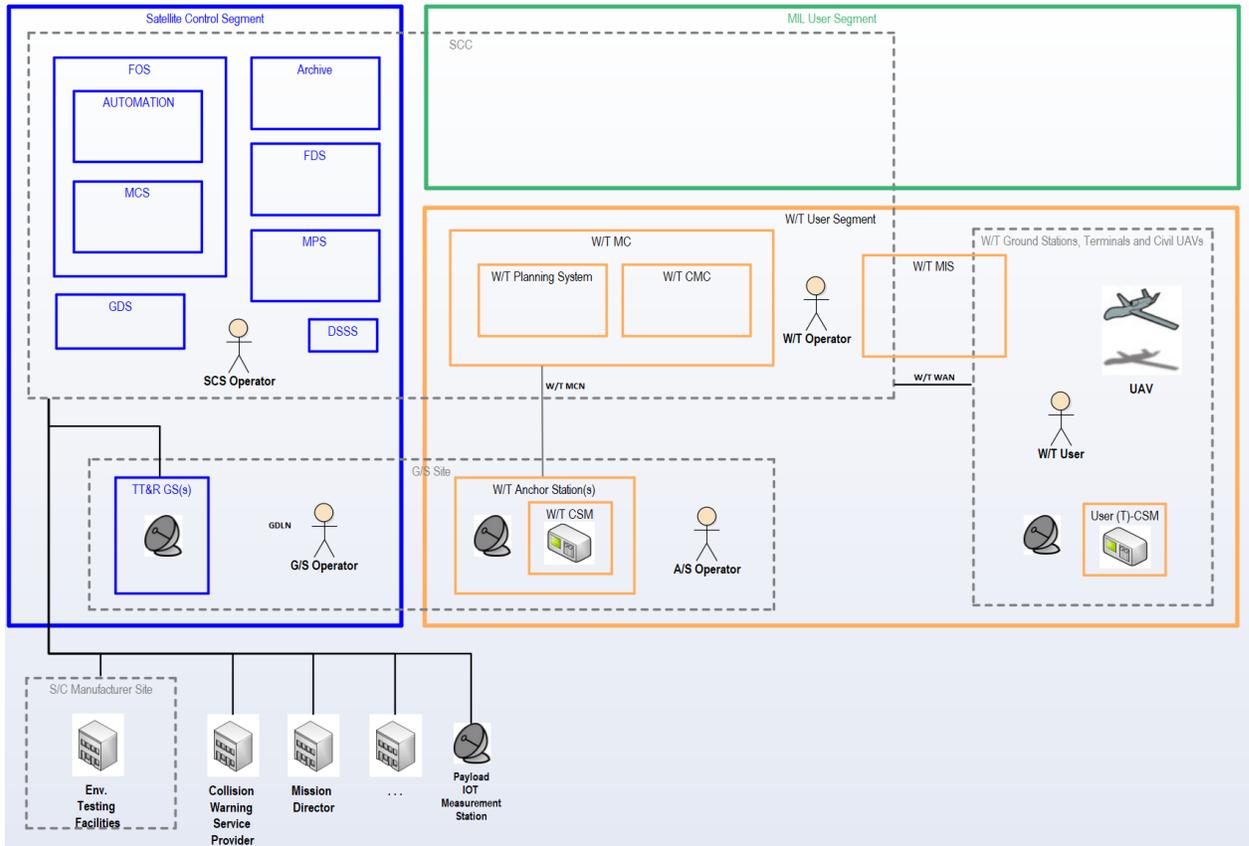

**Figure 1: H2Sat Ground Segment Overview**

The Heinrich Hertz Mission is shared by a large set of different users. As a result, ensuring safe data separation on ground becomes a critical element of any ground segment software solution to facilitate secure and reliable usage across diverse user groups. The concept of Data Separation in the H2Sat Mission is primarily driven by two types of access limitations:

- Open-Restricted: Telemetry of the military payload is restricted and is under no circumstances allowed to enter the W/T User Segment.
- Need-to-see Principle: Even though all scientific/technical experiments are non-restricted, experiment data (i.e., telemetry) should only be accessible to the respective experimenter.

## 2.1 Open-Restricted Data Separation

The open-restricted separation is a great challenge for the entire system. As described in Section 2, the complete Ground Segment logically consists of two segments, the Satellite Control Segment (SCS) and the W/T User Segment. Only the SCS is restricted (as VS-NfD) while the W/T User Segment is non-restricted. This means that the security gradient between open and classified data lies within the system. This imposes several challenges to ensure that no restricted data enters the W/T User Segment which are solved using several redundant security mechanisms.

The spacecraft MIB contains for each telemetry parameter its security classification. Each telemetry parameter can be either open (unclassified) or restricted (classified/VS-NfD). Based on this classification, all data is automatically sorted by the central M&C component into a hierarchical tree data structure. VS-NfD data is stored under a dedicated node data.restricted while open data is stored under the node data.open. Below these nodes a separation in unprocessed (TM received from the spacecraft or ground segment equipment) and derived (for derived parameters/synthetics) is made. The deeper structure of the next levels is not of interest for the discussion about telemetry splitting. The tree structure represents a high-level separation of VS-NfD data from open data. Only data originating from the data.open node is allowed to enter the non-restricted W/T User Segment. This is realized by a dedicated TM Splitter component that all telemetry passes through before it enters the W/T User Segment. Similar mechanisms are used for ground and space event messages, e.g. PUS 5 on-board events or ground power outage notifications. In addition to this, a dedicated packet filter is employed to ensure that only packages carrying unclassified information are allowed to leave the Satellite Control Segment and enter the W/T User Segment, which happens on network level.

*2.2 Need-To-See Data Separation*

The realization of the need-to-see principle benefits from the requirement of the open-restricted separation. Parts of the technical solution for the open-restricted separation can be reused to also implement the need-to-see principle. The general idea of the need-to-see principle is that data is only available to "authorized" clients. While this applies in general to different aspects of the system, this section covers the separation of (experiment) data between different experimenters. No experimenter shall have access to experiment results from a different experimenter.

The special design of the H2Sat payload complicates this problem. While the H2Sat mission is equipped with 13 communication experiments and 9 technology demonstrations, those experiments do *not* come as separated on-board components, but rather are derived as configurations from the single complex scientific/technical payload. This payload consists of several antennas, MPMs, filters, etc. and different experiments are in principle formed and defined by routing signals differently through the payload.

The consequence is that it is not possible to assign a certain on-board component to a certain experimenter which would allow for easy access management. Instead, an MPM might be used both by experimenter A and experimenter B for their experiments. This means that the telemetry generated by the MPM belongs to experimenter A during execution of experiment A, to experimenter B during execution of experiment B and to neither of them during any other time. Therefore, time-based access restriction is necessary to fulfil the need-to-see principle.

Separation must be performed both in term of *content* and *time*: *Content Separation* refers to the definition what telemetry (as defined in the spacecraft MIB) is accessible for which user group. *Temporal Separation* refers to the fact that some experiments share components (for example on their signal path) such that telemetry of shared component shall only be accessible when corresponding Activity Requests of that user group exist.

All telemetry in the H2Sat Ground Segment is centrally handled in the tree structure containing any parameter in the H2Sat Ground Segment, space or ground. No W/T User Group has access to this master copy. Instead, for each experiment, a dedicated sub-tree is generated consisting of a copy of each relevant TM parameter for this particular experiment. For this, a DataProcessor is configured for each experiment that collects any relevant raw and derived parameter for the experiment and copies it to the respective subtree. For each subtree, access control is defined providing access to the subtree for the respective W/T User, only. Any HMI component as well as the telemetry stream of a given W/T User can only connect to his corresponding subtree. This way, a W/T User can only access telemetry from his respective subtree as created via the corresponding DataProcessor. This mechanism ensures Content Separation and is shown in Figure 2.

To ensure Temporal Separation, the DataProcessor is time controlled. By default, it is switched off and therefore not copying any TM parameters to its subtree. At the beginning of each W/T Activity Request, a dedicated GOP is executed that switches on the DataProcessor to start copying telemetry to the W/T User's subtree. At this point of time, the W/T User can see deferred telemetry in his HMIs and receives deferred telemetry via the W/T Telemetry Stream. At the end of each W/T Activity Request, the DataProcessor is switched off by calling a dedicated GOP. Now, the W/T User cannot access the deferred telemetry anymore. This way, each W/T User tree is populated only during the

Internal

execution times of the respective experiment. In the Mission Archive, the W/T User is given access also only to his corresponding user tree. This mechanism ensures Temporal Separation, both for deferred and archived telemetry.

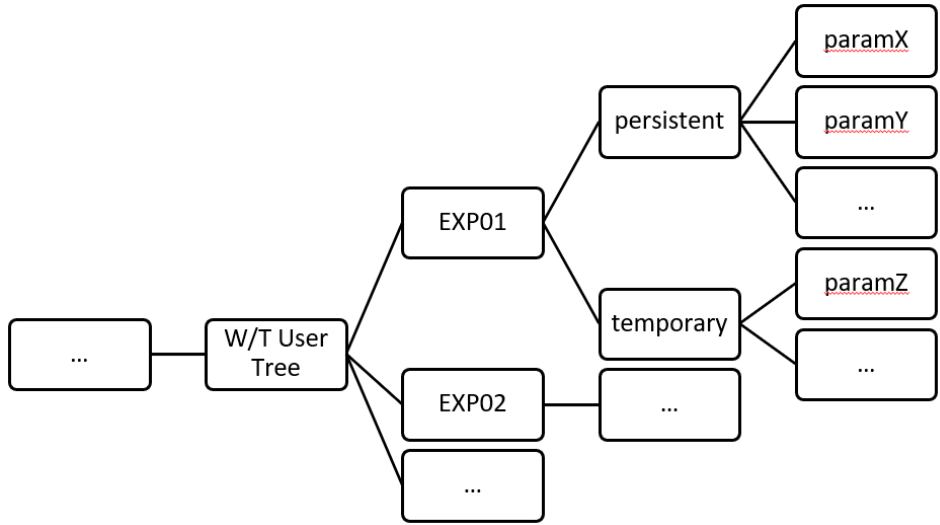

**Figure 2: Tree Layout for a mixed-mode DataProcessor**

The mechanism of switching on/off DataProcessors realizes Temporal Separation by recording all or none of the telemetry parameters corresponding to an experiment. It might be the case, however, that an experiment consists of shared components whose telemetry should not be accessible all the time to a W/T User and of proprietary components whose telemetry should be available always and only to this W/T User. For these kinds of situations, a *mixed-mode DataProcessor* can be used. Instead of being switched on/off completely, the DataProcessor copies one part of the telemetry (corresponding to the proprietary components) all the time. Telemetry of shared components is only copied when commanded so by receiving a command via the respective GOP. In this case, the subtree of the W/T User looks like as depicted in Figure 2. Telemetry corresponding to proprietary components is copied to the subtree "persistent" while telemetry corresponding to shared components is copied to the subtree "temporary".

In summary, this data separation mechanisms give us the possibility to automatically provide external users access to certain telemetry without violating any security principles like need-to-see or possibly leaking restricted data. Data separation is controlled by the same GOPs that are used by operations staff and external users to trigger the communication experiments on board. By employing several mechanisms that ensure data separation, we are confident in the system to provide a secure, external access to support lights-out operations, and provide a self-service portal for external scientific users.

### 3. *Lights-Out Operations*

Humans have been launching and operating satellites since the 1950s. In those early days, operations relied on heavy manual controls and rudimentary data processing methods, but they established the foundational practices of satellite monitoring and communication.

Over the decades, advances in computer technology and automation transformed the design of satellite missions. Ground control systems evolved from hands-on, human-operated control stations into sophisticated ground data networks. These improvements not only expanded our capabilities in satellite technology, in terms of communication, navigation, and Earth observation but also set the stage for a paradigm shift in operational efficiency on the ground.

The ground segment for Heinrich Hertz builds on a flight-proven, manually operated software core to allow manual control during special operations (like LEOP or critical manoeuvres). However, the main mode of operations is a fully automated one: **Lights-Out**. While the satellite control centre is fully staffed during office hours (i.e. Mon – Fri, 9 – 5), satellite operations during more arduous and thus expensive times are flown completely automated by the system. This section will focus on the technology and operations principles used in the H2Sat ground segment for this.

Internal

The H2Sat Ground Segment Software System was designed and realized by using building blocks from the CGI COTS "Pleniter®" suite [3]. This software suite provides components for all aspects of a ground segment. The available components are:

- Pleniter Control – The central M&C facility responsible for monitoring and commanding both space and ground assets
- Pleniter Automate – This component provides means to automate any activity in the ground segment and on-board via procedure definition and execution
- Pleniter Plan – The central planning facility responsible for creation and execution of a conflict-free overall mission schedule
- Pleniter Orbit – A component for all aspects concerning flight dynamics, e.g., manoeuvre generation
- Pleniter Store – The central facility to store and distribute any data originating from space and ground

Although the components are all available individually and are based on flight-proven CGI software that has been in use for years in other missions, e.g. Galileo or CMCF, their deployment in H2Sat is the first time they have been used for all aspects of a satellite mission. While they each individually can be connected to other systems, e.g. the Plan module can work with a different MCS or FD system, they work best together as all software is maintained by one company and was optimized to work together.

H2Sat is a complex mission with many external users. The military part of the mission is solely used by the German Armed Forces. The scientific/technical payload of the mission, however, consists of 13 communication experiments and 9 technology demonstrations. Operating such a complex payload with a relatively small team requires a high level of automation.

The focus of the automation concept lies on supporting the operational concept that foresees operation only during normal business hours on workdays. Outside of office hours, only on-call personnel are available. Therefore, nominal operations that takes place outside of office hours must be ensured without any human interaction. Furthermore, automation needs to cover and identify anomaly situations and must be able to take appropriate actions, for example automatically notifying the on-call operator. As with any automated system, careful consideration must be taken to find the correct balance between automation and responsible operation. This does not so much relate to a risk to lose the complete mission as the satellite design ensures that the satellite can operate autonomously for 48h. Therefore, any automation of the system can be switched off to ensure full manual control in contingency situations.

### 3.1 Monitoring and automatic notification

The Monitoring and Control System (Pleniter Control) constantly monitors and processes any parameter originating from space and ground. A various number of processing functions is available for statistical analysis of telemetry such as statistical computations, rule-based computations and trend analyses. The computed results can itself be taken as input for further processing steps which allows the definition of arbitrary processing pipelines. These processing can either result in a derived parameter/synthetic which is visible to operators and archived, or alternatively can become an event message in the system such that automatic reactions to this event can be configured and triggered.

This processing is used to monitor all parts of the system. Because Pleniter processes both ground and space telemetry in the same way, definitions for out-of-limit constraints and other anomalous events can be used for both space anomalies and ground anomalies or even combine the data to look for space-ground anomalies. For example, an out-of-limit situation could easily be defined in the system for the situation that differing power levels between ground and space occur that are unexpected. This unified, holistic approach enables the operations team themselves to further define constraints and limits, even during the mission, within minutes to allow for continuous improvement of the system and its monitoring and anomaly detection capabilities. We learned during commissioning of the lights-out mode that this continuous improvement with quick turn-around is required for safe and successful lights-out operations as there were some anomalies where the system didn't behave as expected, which was quickly resolved on the next working day by the operations staff.

As a baseline (that is still operator-configurable) per default all out-of-limits definitions from the MIB are automatically read and applied to avoid human error. The operators still have the option to add new limits and

Internal

constraints or (in direct communication with the satellite support provider), disable certain limits as they are currently expected. None of these actions require software changes or software support team intervention.

When an event message is distributed throughout the system components (starting from the component that was configured to trigger it), the system can be configured by the operations team to trigger appropriate responses. These responses can take different forms:

- HMI Alert Message
- SMS Notification
- Automated Procedure Execution

The easiest response that is enabled by default is a visible alarm in the system. This is automatically shown to all operators that are in the control centre at the time. This is mainly relevant for office hours operations and remote anomaly investigation.

A second form of notification that can be configured is an automatic SMS notification using different cellular network providers. We deployed four SMS gateways that take TCP connections from the system and send out SMS using LTE cell towers. Two SMS gateways are deployed per control centre for local redundancy. Additionally, two SMS gateways are deployed in the remote, dislocated control centre to support notifications in case of catastrophic emergencies like flooding, total power outage including UPS failure and other control centre destroying events. All SMS gateways use different German cellular networks to increase the redundancy and mitigate a potential risk of a service interruption or outage of a particular cellular network.

On the staffing side, at least two people are always notified outside of office hours of an anomaly to make sure that all events are read, and appropriate actions are taken. The operator can then, from remote, perform an initial assessment of the situation. For this, we provide the operator with the "Operator Remote Access" interface (ORA) to investigate all anomalies pertaining to unrestricted data and event notifications. This interface is based on the external user access described in Section 4, but was tailored to the specific lights-out use case. In detail, this allows the remote operator to act as a remote "super experimenter", granting them rights as if they were any experimenter. This allows them to investigate unrestricted anomalies while still following the mission's security rules and requirements. An example display is shown in Figure 3. If the anomaly is not a false alarm, or if the remote, unrestricted-only data access isn't enough to investigate the anomaly, the operator will have to come in to the satellite control centre to further investigate and resolve the situation on-site.

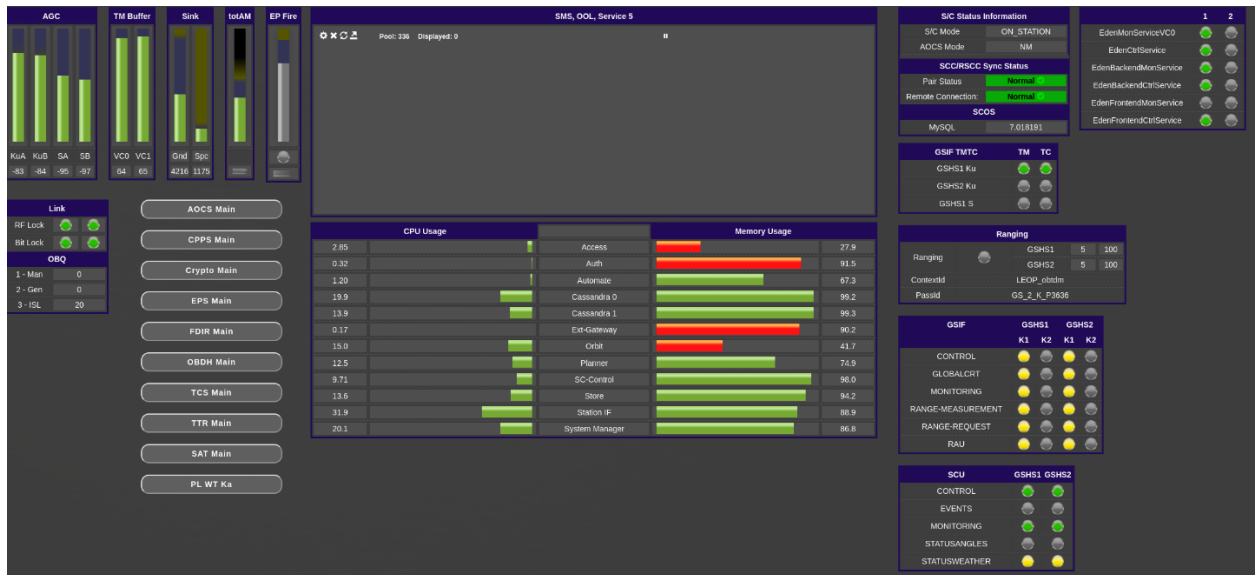

**Figure 3: Remote quick-checkout display, giving operators an overview of the situation in the control centre.**

The third option to react to an anomaly event (Automated Procedure Execution) is described in the following section.

Internal

### 3.2 Automatic commanding

The H2Sat ground segment is equipped with the Pleniter Automate module which acts as the bridge between the high-level planning services provided by the Mission Planning System (using the Pleniter Plan module) and the lower-level control of the Monitoring and Control System (powered by Pleniter Control). It consists of a procedure execution engine for automatic control of schedule and procedures and provides displays to monitor and control operations at a higher level than is provided by the MCS.

Schedule and procedure execution tightly integrate with the MPS and MCS. A web-based editor is used to set up mission definitions (e.g. SCOS Mission Information Base, MIB) during the operations preparations and routine operations phase. Execution engines (e.g. schedule execution engine) fulfil the actual function of the asset which is monitored and controlled using GUI displays during the operations phase. Finally, the history of the asset's function is archived for later evaluation.

The Automation framework consists of the two Pleniter components Procedure Execution Engine (part of Pleniter Automate) and Schedule Execution Engine (part of Pleniter Plan). Both operate on different levels: The Procedure Execution Engine processes procedures while the Schedule Execution Engine processes schedules. See also Figure 4 for an example.

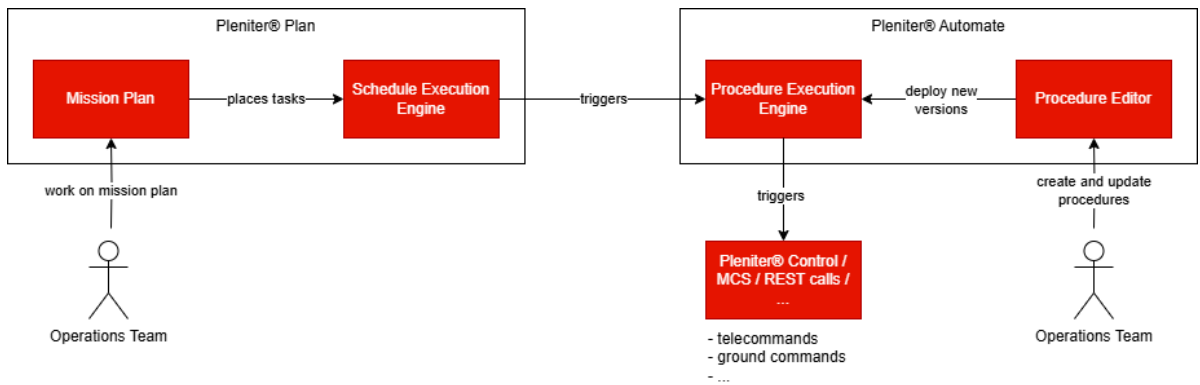

**Figure 4: Explanation of cooperation between Schedule Execution and Procedure Execution**

These modules work together in the lights-out mode to complete a much more powerful feature than simple notification of an operator: Automatic telecommand and ground command responses to incidents. In our ground segment, the operations team works together with the satellite support team provided by the satellite manufacturer to define appropriate and validated response procedures in the graphical procedure editor which are deployed to the operational system and configured to be accessible to the automated response feature in the lights-out mode. This means that any telecommand or ground command that is available in the system (meaning all control that the operations team also has manually) is available as a scripted, well visualized and most importantly validated operational procedure that will be triggered fully automatically by the system in case of a defined anomaly. As described above in Section 3.1, this can be an out-of-limit event that is detected on ground or an on-board satellite event that was downlinked in a PUS 5 event. Of course, this also can be a ground event originating in the MCS or FDS systems, like TM loss or similar ground mission event or the finished computation of certain FD parameters.

These procedures are placed in the central procedure store. If a configured event occurs, the automatic response begins by automatically requesting the execution of the procedure in the Pleniter Plan module. The use of the MPS instead of direct execution via MCS allows for more safety in the execution by allowing constraint checks to execute before the procedure execution. From a technical point of view however, also triggering the MCS to perform tasks, e.g. telecommands is possible. The MPS will trigger execution of the procedure using the Procedure Execution Engine. If during this process any error occurs, this automatically becomes a notification to the on-call personnel to make sure that the appropriate response to the event is triggered, even if the automated systems should fail. More details about automated schedule generation, constraint resolution and procedure execution will follow in Section 4 when we discuss automated user activity request and automated user schedule deconfliction.

In summary, this means that based on telemetry data or statistical analysis, automatic responses can be configured. These responses range from automatic notification of operators via the HMI, automatic notification of on-call operators to issuing of automatic commands to ground and space. Several examples of automated workflows, that involve the lights-out mode and are also needed to allow external self-service user access, are described in Section 5.

## 4. External User Access

Both payloads, the scientific/technical as well as the military payload, are operated by the respective external user(s). External users are not located in the control centres but operate their payload from remote. This might, for example, be an academic institute or university premises for the scientific/technical part of the mission.

External User Access in the H2Sat Ground Segment is designed for a maximum level of automation and (user) autonomy to free the operator from routine tasks. For nominal operations of their experiments, the W/T User needs none or only little operator support.

To enable remote operation for external users, the H2Sat Ground Segment Software System provides them a secured web interface. External users can fully operate their payloads via this interface (for nominal situations). This means:

- Access to the online telemetry related to their payload
- Access to archived telemetry related to their payload
- Commanding capabilities for their payload

For telemetry access, both online and archived, the external users have access to the same set of telemetry displays that are available for the operators inside the control centres. Pleniter allows operators to quickly and flexibly design and deploy web-based displays by either using pre-defined widgets or by writing HTML themselves which enables the subsystem experts to autonomously create very specific and highly complex displays. Due to the reuse of the same displaying technology as is used inside the control centres, duplicate work is avoided. Telemetry can be displayed as single values (engineering or raw), on tables or on graphical charts. Historical telemetry can be retrieved, either for immediate statistical analyses inside the web interface or downloaded for further offline detailed analysis with external tools. This way, experimenters can follow their experiment "live" during experiment execution and/or download the full set of telemetry data after completion of an experiment activity. An example display demonstrating the capabilities of the flexible web HMI is shown in Figure 5.

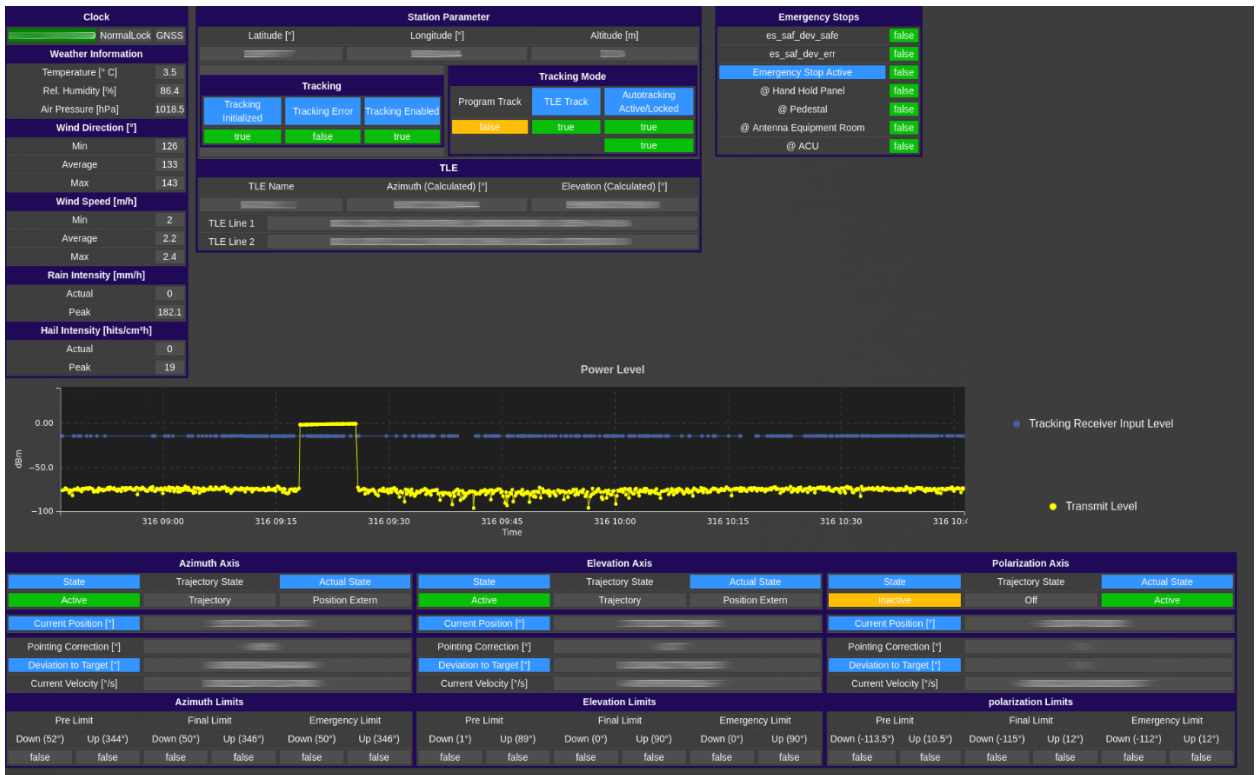

**Figure 5: Example operator-created HMI fed by live satellite and ground telemetry data**

Internal

### 4.1 External command capabilities

External users have full access to a pre-defined set of commanding capabilities of their payload. An external user can command their payload (during nominal situations) without any interaction of the operating personnel in the control centre. The user's planning requests -called *Activity Requests (AR)*- are automatically checked by the system for correctness and constraints. A rich set of different constraint types are available in the Mission Planning System modelling, for example, inter-task dependencies, on-board resource usage and maintenance slots. The Mission Planning System automatically generates conflict-free schedules (if possible) and forwards them to the Automation component for execution. This is used heavily in the W/T segment to make sure no two conflicting experiments can run at the same time. If such a case occurs, rules define which tasks takes priority or, alternatively, if the conflict is resolved using first-come-first-serve principle.

Planning sessions occur fully automated on a regular basis. The time between issuing a planning request and execution of the requested task (give that the request did not cause any non-solvable conflicts) is usually less than 10 minutes. This means that not only very short-term experiments can be scheduled and executed but also that changes to running experiments are possible. The concept to change a running experiment is called *User Execution Request (UER)*. UERs enable the W/T user to not only rely on fully pre-scripted experiments but to also react on certain circumstances during the experiment execution.

Pleniter Plan can accept planning requests up to a configurable time point into the future ("Planning Horizon"). There are no practical limits for the planning horizon, i.e., it could be 1 year into the future. The definition of a planning horizon refers to scheduled activities that need a planning session including constraint (and resource) checks. Direct commanding via the Procedure Execution Engine needs even less lead time.

The set of ARs and UERs is pre-defined by the control centre (i.e., W/T Mission Control) by the operations team in cooperation with payload experts. This ensures that commanding capability of external users is restricted to a well-understood subset of the payload functionality. AR and UER definition are done in collaboration with the experimenters and can easily be changed during the mission. The operators can define, using the Automate module, graphical procedures that map the needs of the experimenter into telecommands/telemetry checks and make this procedure available to the external user. The option of adding new user experiments (using, of course, the same physical on-board payload) during the mission is also flexibly available to the operations team.

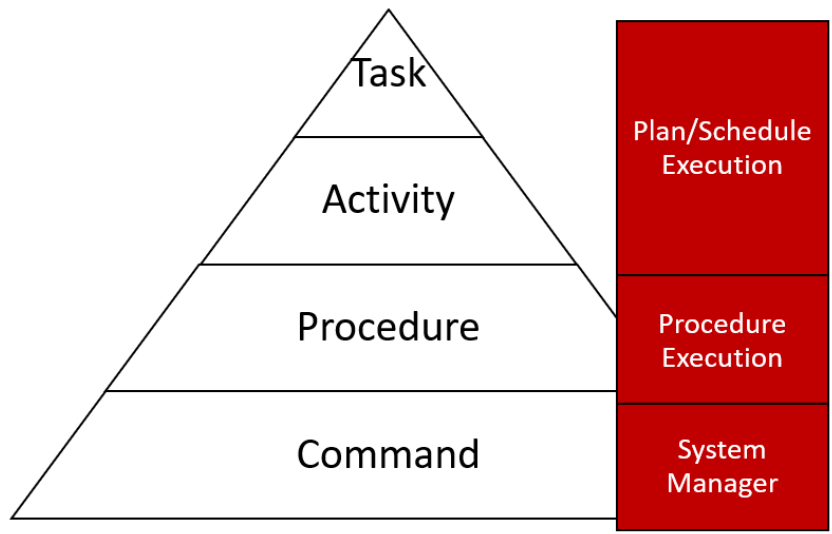

**Figure 6: Hierarchy of commanding concepts in Pleniter**

As you can see in Figure 6 above, there are several concepts in Pleniter about "commanding the satellite". These become more abstract and safer to the top and more manual towards the bottom.

*Commands* reside on the lowest level and refer to both spacecraft telecommands as defined in the spacecraft MIB and commands targeting any ground equipment. Commands and are processed and issued by the Pleniter System

Internal

Manager and are usually not directly planned/executed by any user (except for the LEOP phase or contingency situations).

*Procedures* wrap telecommands and are defined with the Procedure Editor, a GUI that is part of Automate, and are executed by the Procedure Execution Server. Procedures are validated via the mission Flight Operation Procedures (FOPs) targeting the spacecraft and the Ground Operation Procedures (GOPs) for ground equipment. Procedures can be executed directly via the Procedure Execution Server when immediate commanding is necessary. This circumvents any resource checks and other conflict checks because constraint checking is done on the next level, i.e., for Activities. Procedures form the smallest commanding building block operationally used in H2Sat. Note that procedures can contain any kinds of control structures (such as loops and conditional branching) within their body. Procedures can have parameters which can be forwarded to the encapsulated telecommands.

*Activities* do not live on their own but constitute tasks. Activities can only be executed via triggering their encapsulating task, not individually. An activity wraps a single procedure. Activities can have parameters which are forwarded to the encapsulated procedure.

*Tasks* refer to a sequence of (potentially overlapping) activities. Note that there are no control structures on task-level, i.e., tasks are fully described by their sequence of activities. Tasks can have parameters which are forwarded to the constituting activities. The H2Sat concepts of Activity Requests and User Execution Requests map to the concept of a task.

The external user can request the execution of a task (the set of tasks they can choose from is limited using access control lists) freely to any time slot. The system will then check various configured constraints like for example power levels, exclusion constraints between different experiments, etc. Constraints can also be added during flight by the operators when new needs arise. Then, after all checks are done the system notifies the user if their request was placed.

The amount of details the external users see of Tasks of other stakeholders (e.g. other experimenters or even platform/payload activities) can be configured and varies between full details (this is configured for operators so they can see all details of ongoing payload activities), anonymized tasks (the experimenters are only shown that other experimenters use the payload during certain times) to no access at all, such that they only get informed about the successful placement or rejection of their activity once the automated checks are run and their request is confirmed or denied. This is configurable on user- or group level using access control lists.

In summary, this concept gives the external user a lot of flexibility to schedule and perform the communication experiments they are responsible for. This "self-service" HMI allows the operator to be very independent of the operations staff, allowing them to focus on platform-centric tasks and allowing tighter budgets while still giving users the flexibility they need to perform GEO communication experiments.

### *4.2 Measuring equipment*

In addition to a web interface for the above-mentioned purposed, the H2Sat Ground Segment also provided a set of measurement equipment that can be used by the W/T User to perform measurements on the user link. Measurement equipment includes, for example, a vector network analyser, a spectrum analyser and other RF Test Equipment. By providing the measurement equipment centrally on the anchor stations, any experimenter can benefit (if required) from these devices without the need to purchase them on their own.

To give the W/T User access to these shared devices, a dedicated access mechanism is implemented in the Anchor Stations which is triggered remotely as part of the experiment activation procedure, meaning access is automatically granted when an user experiment begins. In addition to accessing shared equipment, experimenters are also allowed to equip the Anchor Station with own measurement devices.

### 4.3 Experience and Experimenter Feedback

Since H2Sat finished its LEOP and IOT phases in 2023 and entered its routine operations, several communication experiments have been conducted. The positive feedback received from all of them have been instrumental to finetune the W/T User HMI. One of them is the GeReLEO experiment [4] which relays information from a LEO satellite through H2Sat to ground, as H2Sat has a much longer contact time with the LEO satellite and the ground station.

When performing this experiment using the external HMI, the user feedback was great, highlighting the useability improvements of seeing live satellite telemetry during the experiment execution via graph visualisations and telemetry tables. Their main remark was especially that the operations team was able to reproduce layouts for the displays that were similar to the graphics they used in the payload user manual.

Internal

W/T payload engagement on improvements was a key indicator of their satisfaction. Enhancements like the desire to follow the telecommands sent to the satellite was received and as part of our continuous improvement process, we are currently investigating options how to realize this without violating security principles of the mission.

To summarize, the ground and space segment of H2Sat work together to provide external experimenters a 24/7 accessible GEO platform for their communication experiments. It should be noted that due to the high degree of automation and the high configurability of the ground segment, the lights-out mode and external self-service are run in parallel, meaning that even in the staff-reduced lights-out mode the experimenters self-service HMI is available and can be used normally.

## 5. Shared Mechanisms

There are several automated mechanisms that are important to both aspects of this paper, lights-out and the external user access:

### 5.1 CSM Monitoring

CSM monitoring refers to the following tow tasks:

1. Monitoring power limits imposed by frequency coordination and ITU regulations
2. Intruder detection and interference detection

For this, a dedicated component W/T CSM was designed for the W/T User Segment that fulfils both purposes. It processes CSM measurements as received from dedicated RF Test Equipment located in the Anchor Station. The purpose of the W/T CSM components is to perform the necessary calculations for power limit monitoring and intruder detection and to raise events in case of identified findings.

### 5.2 Monitoring Power Limits

Power Limit Monitoring is a completely automatic process performed for each currently executing W/T Activity Request that executes a telecommunication experiment. Activity Requests are executed by the Procedure Execution Engine located in the Satellite Control Segment. Upon execution, a dedicated GOP automatically triggers W/T CSM before any communication link is established. Its purpose is to configure W/T CSM for correct power limit monitoring. For this task, the GOP passes the relevant parameters (like instance id of the current Activity Request, the frequency, the bandwidth, footprint losses, etc.) to W/T CSM and triggers the start of power limit monitoring. If W/T CSM detects a violation of power limits, it raises an alarm to the user interface of the console in W/T Mission Control resulting in an appropriate message displayed to the operator. In addition, the event is archived in the W/T Mission Archive. If deemed necessary, it is also possible to execute automatic commands based on CSM measurements, for example to mute all W/T transponders in case of limit violations.

A second dedicated GOP is scheduled at the end of the Activity Request. It notifies W/T CSM that the experiment has ended, and that no user signal should be present anymore. W/T CSM will then continue monitoring limit violations for a configurable time span (e.g.: 10 minutes) to allow the experimenter to shut down his equipment. If after the time span still a user signal is detected, W/T CSM raises an alarm via the CSM Events interface to W/T CMC which is then processed as described above. In addition, an alarm to the operators can be configured.

### 5.3 Intruder Detection

Intruder Detection operates in two modes: during and outside of experiments. The default mode of intruder detection assumes that no signal should be present in any channel. If a signal is nevertheless detected, an alarm is raised via the CSM Events interface to W/T CMC. W/T CMC ensures that an appropriate message is displayed to the operator and triggers archiving of the alarm. This default mode of intruder detection is assumed to be running all the time. It can be switched on and off via the CSM Commanding interface from the W/T CMC. If a W/T Activity Request is running, the channel(s) used by this experiment are not empty anymore. W/T CSM is informed via a dedicated GOP scheduled at the start of the Activity Request what bands are expected to be used by the experiment and ignores these during the runtime of the experiment. In case of a detected intruder signal, a dedicated alarm is raised and processed via W/T CSM as described above.

## 6. Conclusions

The H2Sat Ground Segment is designed to support an operational concept with a reduced staff size based on the Pleniter® suite [3]. The main drivers to achieve significant workload reduction for the operational crew are a high degree in automation for monitoring and control of the space and ground segments as well as a flexible and highly autonomous access for external users which frees the operator from many repetitive service desk tasks. External users can monitor and control their payloads via a web interface without the need to directly interact with operators. Nominal experiment planning and execution is performed fully automated without the need for human interaction. Practical experience with external academic users (e.g. GeReLEO experiment [4]) shows warm reception of this capability. With a broad set of automated functions, monitoring and control of both space and ground assets can be performed with a reduced staffing size for the Heinrich Hertz mission.

## Acknowledgements

*CGI Deutschland B.V. & Co. KG has been awarded the contracts for the realization of the ground segment (Phase C/D) and the operations (Phase E1) from mission preparation, LEOP, IOT and routine operations of the German national satellite communications mission Heinrich Hertz (H2Sat). Heinrich Hertz is conducted by the German Space Agency at DLR, with funding from the German Federal Ministry for Economic Affairs and Climate Action (BMWK), and in collaboration with the German Federal Ministry of Defence (BMVg). OHB-System AG, as the prime contractor, was commissioned to develop and build the satellite and to conduct the phases C/D and E1 of the Heinrich Hertz mission operations.*

Internal